\documentclass[sigconf,nonacm]{acmart}  
\usepackage{multicol}
\usepackage{multirow}
\usepackage{array}
\usepackage{colortbl}
\usepackage{arydshln}
\usepackage{bm}
\usepackage{mathtools}
\usepackage{url}

\AtBeginDocument{%
  }

\begin{document}

\title{Does Listening Matter? Backchanneling and Nodding in AI Clone}

\author{Koji Inoue}
\email{inoue@sap.ist.i.kyoto-u.ac.jp}
\affiliation{%
  \institution{Kyoto University}
  \city{Kyoto}
  \country{Japan}
}

\author{Kazushi Kato}
\email{katou@sap.ist.i.kyoto-u.ac.jp}
\affiliation{%
  \institution{Kyoto University}
  \city{Kyoto}
  \country{Japan}
}

\author{Tatsuya Kawahara}
\email{kawahara@i.kyoto-u.ac.jp}
\affiliation{%
  \institution{Kyoto University}
  \city{Kyoto}
  \country{Japan}
}

\author{Shunichi Kasahara}
\email{kasahara@csl.sony.co.jp}
\affiliation{%
  \institution{Sony Computer Science Laboratories}
  \city{Tokyo}
  \country{Japan}
}

\renewcommand{\shortauthors}{Inoue et al.}

\begin{abstract}
AI clones that imitate a specific person typically reproduce what the person says and how they sound, but not how they listen.
We investigate whether adding multimodal listening behaviors gives such a clone more presence and authenticity.
We integrated verbal backchannels and head nodding, driven by real-time prediction models, into an AI clone equipped with voice cloning and LLM-based responses.
In a within-subjects study (N=35), adding these behaviors significantly improved the perceived attentiveness of the avatar, the sense of talking with the real person, and the feeling of co-presence.
These results indicate that AI clone fidelity should extend beyond voice and response content to include interactive listening behavior.
\end{abstract}




\maketitle

\makeatletter
\begingroup
  \renewcommand\thefootnote{}%
  \let\@makefnmark\relax
  \let\@makefntext\noindent
  \footnotetext{This paper has been accepted to the Late-Breaking Results (LBR) track of the 28th International Conference on Multimodal Interaction (ICMI 2026). This is the authors' preprint version.}%
\endgroup
\makeatother


\section{Introduction}

Recent advances in LLMs and speech synthesis have made it increasingly feasible to create AI clones that imitate specific individuals~\cite{shirvani2025talking,shirvani2026cloning,park2026ai}.
Prior work has explored persona agents and self-clones that reproduce a person’s personality, values, speaking style, and voice using prompts, background information, and zero-shot speech synthesis~\cite{lee2025creating,kasahara2026clone,chen2025neural}.
Recent full-duplex speech models further integrate role conditioning, voice control, and low-latency interaction~\cite{roy2026personaplex}, while work on migratable agents suggests that identity consistency across embodiments can affect trust, likability, and social presence~\cite{tejwani2020migratable}.
Together, these studies show rapid progress in reproducing what a person says and how they sound, but leave underexplored how a cloned person behaves as a listener.

\begin{figure}[t]
  \centering
  \includegraphics[width=\linewidth]{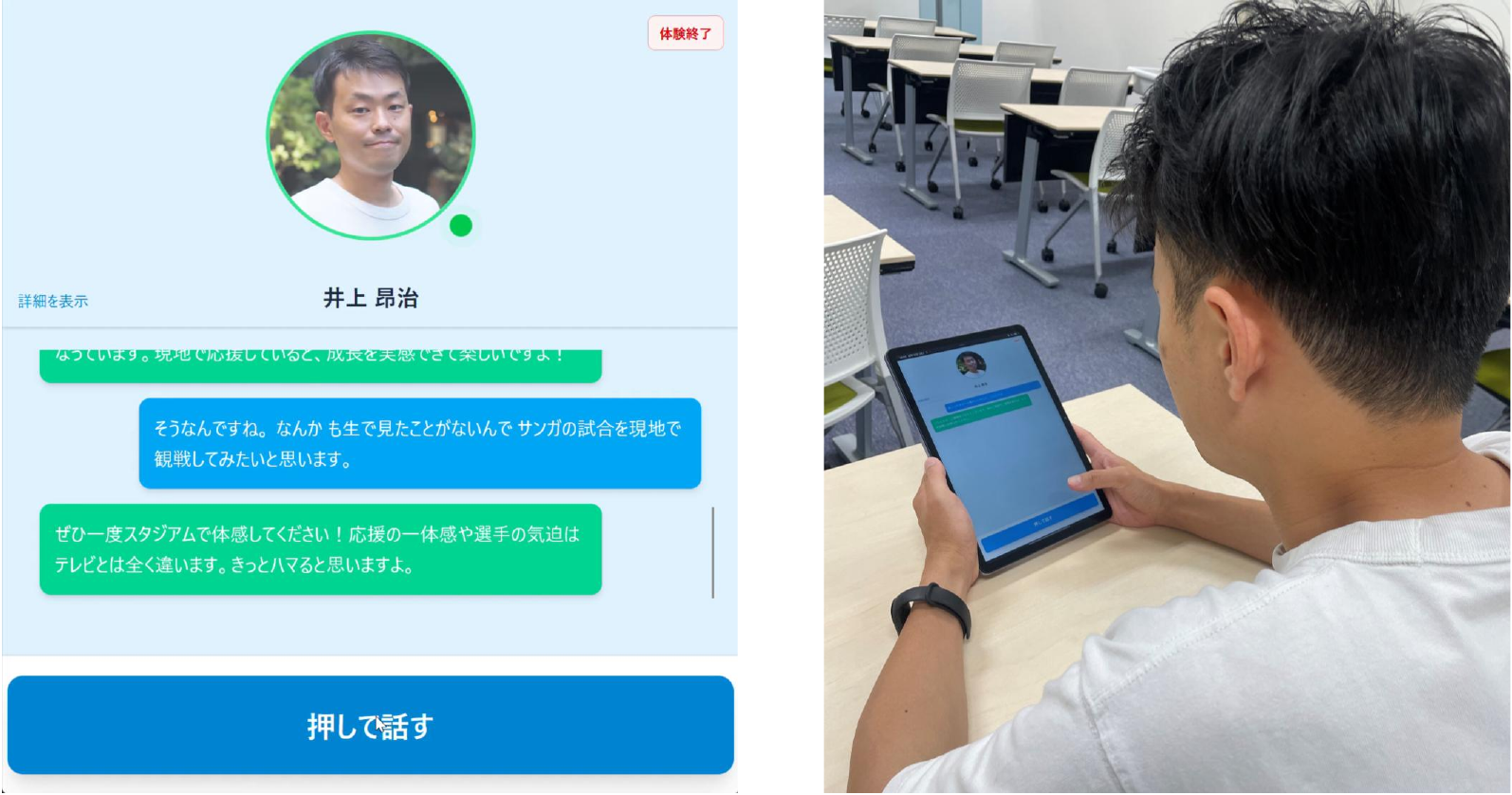}
  \caption{(Left) Interface of AI-clone system (Right) A participant interacting with the AI clone on a tablet during the experiment}\label{fig:interface}
\end{figure}

In human-human interaction, a person's presence is conveyed not only through speaking but also through listening.
Listeners provide brief verbal and nonverbal feedback, such as backchannels and nodding, to signal attention, understanding, and interest~\cite{lin2022duplex}.
Prior work shows that AI agents' backchanneling can function as active listening behavior that enhances user engagement~\cite{jang2024minimal,arjmand2024empathic,jiang2026hear}, while nodding and responsive listener-head motion in virtual agents can improve perceived likability and trust~\cite{cassell1999power,zhou2022responsive,aburumman2022nonverbal}.
These findings suggest that an AI clone's authenticity may depend not only on what it says or how it sounds, but also on how it listens.

In this study, we integrate multimodal listening behaviors, namely verbal backchannels and head nodding, into an AI clone equipped with voice cloning and LLM-based response generation to investigate their effects.
Specifically, backchannels are delivered as short audio cues during the user’s speech, while nodding is visually represented by the simple vertical movement of the avatar's face image (Figure~\ref{fig:interface}).
We conducted an experiment comparing an AI clone with these listening behaviors against a baseline clone without them, evaluating their impact on the user's sense of the original person's presence and the perceived authenticity of the interaction.
The contributions of this study are twofold: first, it expands the concept of AI clone fidelity beyond voice and response content to include interactive listening behaviors; second, it experimentally demonstrates that incorporating backchannels and simple nodding significantly enhances both the perceived sense of interacting with the original person and their overall sense of presence.

\section{System}

\begin{figure}[t]
  \centering
  \includegraphics[width=\linewidth]{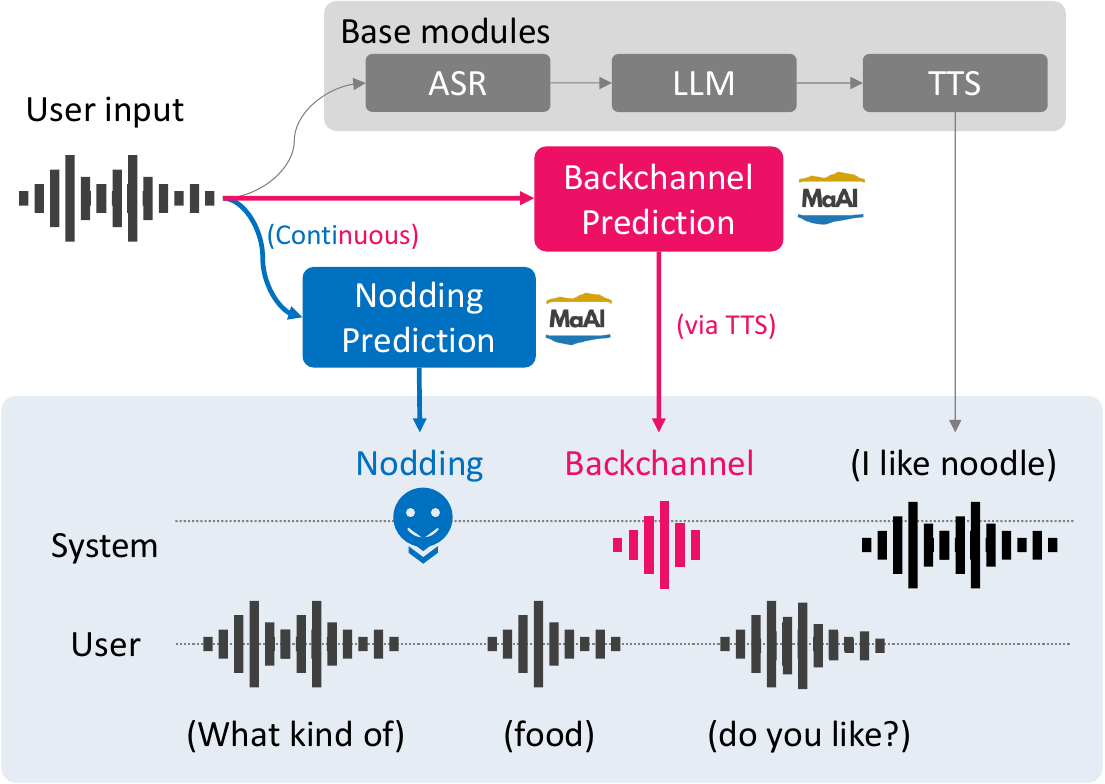}
  \caption{Overview of the system}\label{fig:system}
\end{figure}

The AI clone system used consists of general pipeline modules that include an LLM, along with continuous backchannel and head-nodding generation, as shown in Figure~\ref{fig:system}. 

\subsection{Base modules}

To generate the system's general utterances, the base modules of the system are ASR (automatic speech recognition), LLM, and TTS (text-to-speech synthesis), as utilized from the following cloud services~\cite{kasahara2026clone}:
\begin{itemize}
  \item \textbf{ASR}: Deepgram \texttt{nova-2}\footnote{\url{https://deepgram.com}}
  \item \textbf{LLM}: GPT-4.1\footnote{gpt-4.1-2025-04-14}
  \item \textbf{TTS}: Cartesia \texttt{sonic-3}\footnote{\url{https://cartesia.ai}}, using a voice cloned from the original person
\end{itemize}
The system adopts a push-to-talk interface: the user presses and holds a button at the bottom of the screen, and the recorded speech is processed through the ASR$\rightarrow$LLM$\rightarrow$TTS pipeline upon release.
The user's utterances and system responses are displayed on the screen in a chat-style format (Figure~\ref{fig:interface}).
To reproduce the persona of the original person, a system prompt prepared in advance by the original person describing their speaking style, personal profile, and hobbies is passed to the LLM.

\subsection{Backchannel and Nodding Generation}


Unlike conventional rule-based approaches that rely on rigid acoustic thresholds or silence durations, our non-verbal response generation component employs state-of-the-art continuous prediction models based on Voice Activity Projection (VAP)~\cite{kato2025realtimenodding, inoue-etal-2025-yeah}.
By continuously processing the user's ongoing speech, these models dynamically capture subtle conversational dynamics to predict the optimal onset probabilities for backchannels and nodding.
This approach is implemented using MaAI\footnote{\url{https://github.com/MaAI-Kyoto/MaAI}}, an open-source software specifically designed for continuous backchannel and nodding prediction, which enables highly natural, context-aware reaction timing evaluated at 10~Hz.
To prevent unnatural repetition, a 3-second cooldown is applied after each triggered response.
For verbal backchannels, several patterns of typical Japanese reactive tokens, such as ``un'' and ``un-un'', were pre-generated using the same voice clone of the original person.
When a backchannel is triggered, one of these audio clips is randomly selected and played back, ensuring identity consistency with the TTS module while maintaining conversational variety.
Nodding is visually reproduced by vertically animating the avatar's face image on the screen: a single nod is generated with a 70\% probability and a double nod with a 30\% probability.


\section{Experiment}

We conducted a user study to evaluate the effects of the AI clone's backchanneling and nodding behaviors on users' perceptions of the interaction.

\subsection{Condition}

A within-subjects experiment with two conditions: \textbf{With-feedback} (both verbal backchannels and head nodding enabled) and \textbf{Without-feedback} (both disabled) was applied.
To mitigate order effects, participants experienced the two conditions in a counterbalanced AB/BA assignment.
A total of 35 Japanese native speakers (undergraduate/graduate students) participated in the experiment.
Participants received a 500 JPY bookstore gift card as compensation.

In each condition, participants had a dialogue with the AI clone about the first author's research topics and hobbies.
Before starting the experiment, we presented participants with a text-based demographic profile of the first author.
The two dialogues were experienced independently, and participants completed a post-condition questionnaire after each dialogue.
The experiment was administered by the second author, and the first author did not attend the sessions.
The system was launched via a web browser on an Android tablet, which participants held and operated during the dialogue (Figure~\ref{fig:interface}).


\begin{figure*}[t]
  \centering
  \includegraphics[width=\textwidth]{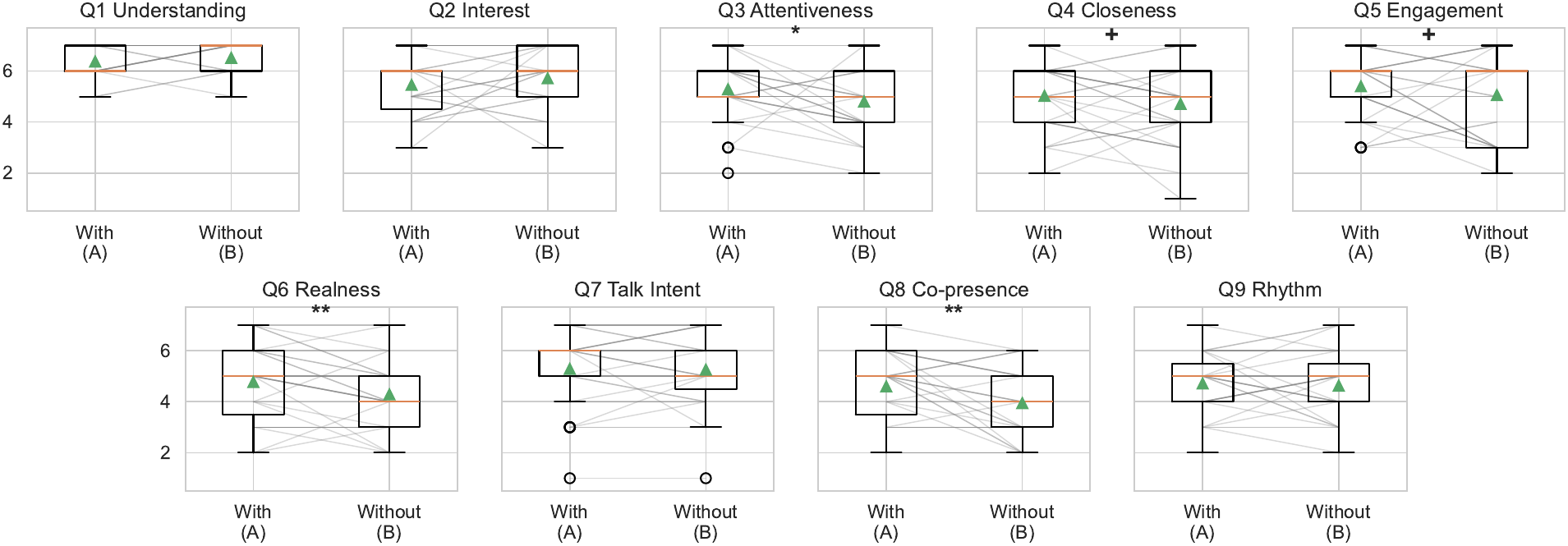}
  \caption{Distribution of ratings for each item (Q1--Q9) under the With-feedback (A) and Without-feedback (B) conditions. Gray lines connect the two ratings of each participant, and triangles indicate the means. ($+\,p<.10$, $*\,p<.05$, $**\,p<.01$, one-sided paired $t$-test)}
  \label{fig:boxplots}
\end{figure*}

Subjective evaluation consisted of nine items rated on a 7-point Likert scale (1: strongly disagree, 7: strongly agree):
\begin{enumerate}
  \item[Q1] (Understanding) I was able to sufficiently understand the content introduced by the avatar.
  \item[Q2] (Interest) Through the dialogue with the avatar, my interest in the introduced content increased.
  \item[Q3] (Attentiveness) I felt that the avatar talked to me attentively while receiving my reactions.
  \item[Q4] (Closeness) Through the dialogue, I felt familiarity and psychological closeness to the person on whom the avatar was modeled.
  \item[Q5] (Engagement) In the dialogue with the avatar, I felt motivated not only to listen but also to return my own opinions and impressions.
  \item[Q6] (Realness) I felt as if I were directly talking with the person on whom the avatar was modeled.
  \item[Q7] (Talk Intent) I wanted to actually talk with the person on whom the avatar was modeled.
  \item[Q8] (Co-presence) I felt a sense of co-presence, as if I were sharing the same space with the dialogue partner.
  \item[Q9] (Rhythm) The interaction timing was smooth, and the conversational rhythm felt comfortable.
\end{enumerate}

In addition to the questionnaire, the system automatically logged each participant's interaction behavior during the dialogue, namely the number of user utterances, their mean length, and the total dialogue duration.
The system also logged the avatar's backchannels and nodding behaviors as system-level logs.

\begin{table}[t]
  \centering
  \caption{Interaction behavior measures as mean (SD)}
  \label{tab:behavior}
  \begin{tabular}{lccl}
    \hline
    \multicolumn{1}{c}{Measure} & With & Without & \multicolumn{1}{c}{$p$} \\
    \hline
    \rowcolor{gray!20}\multicolumn{4}{l}{\textit{User behavior}} \\
    User utterances & 12.89 (2.52) & 13.11 (2.13) & .516 \\
    Mean utterance length (s) & \phantom{0}7.30 (2.91) & \phantom{0}6.93 (3.15) & .195 \\
    Dialogue duration (min) & \phantom{0}4.90 (0.25) & \phantom{0}4.83 (0.19) & .300 \\
    \rowcolor{gray!20}\multicolumn{4}{l}{\textit{System feedback behavior}} \\
    Backchannels & 27.29 (7.55) & --- & --- \\
    Backchannels / utterance & \phantom{0}2.25 (0.92) & --- & --- \\
    Nods & 21.83 (6.10) & --- & --- \\
    Nods / utterance & \phantom{0}1.76 (0.62) & --- & --- \\
    \hline
  \end{tabular}
\end{table}

\begin{figure}[t]
  \centering
  \includegraphics[width=\linewidth]{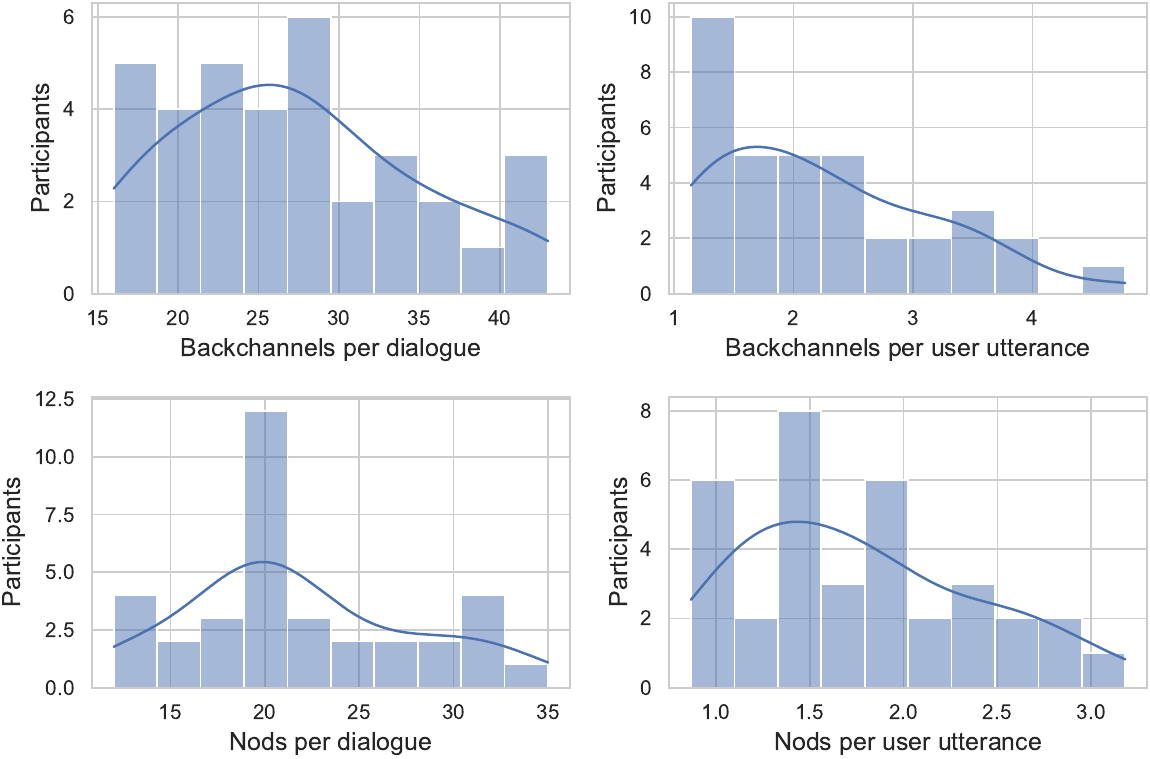}
  \caption{Distribution across participants of the system's listening feedback in the With-feedback condition: backchannels (top) and nods (bottom), as totals per dialogue (left) and per user utterance (right). Curves are kernel density estimates.}\label{fig:freqdist}
\end{figure}

\begin{figure*}[t]
  \centering
  \includegraphics[width=\linewidth]{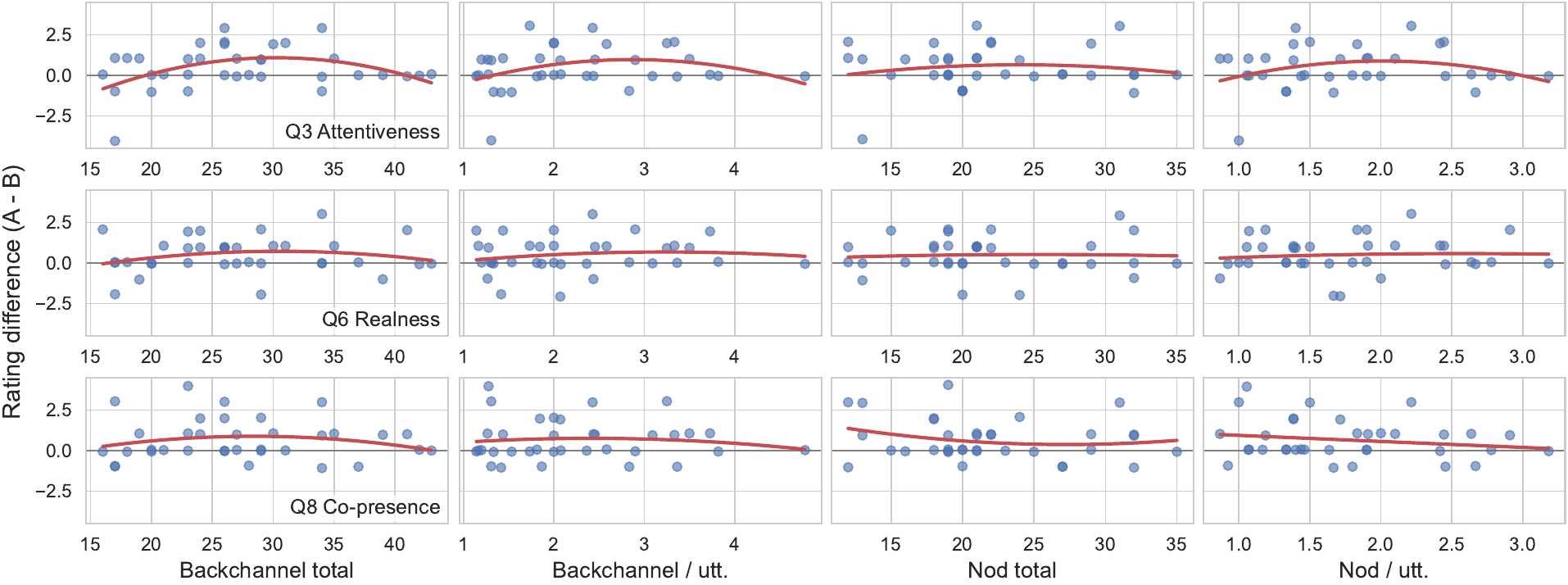}
  \caption{Per-participant rating gain (With-feedback $-$ Without-feedback) as a function of the amount of the system's feedback (backchannels and nods, as total count and per-user-utterance rate) for the three items with a significant condition difference (Q3, Q6, Q8). Red curves are quadratic fits; the horizontal line marks zero gain.}\label{fig:quad}
\end{figure*}

\subsection{Subjective Evaluation}

The nine-item scale showed acceptable internal consistency (Cronbach's $\alpha = .77$ and $.81$ for the With-feedback and Without-feedback conditions, respectively).
Figure~\ref{fig:boxplots} shows the distribution of ratings, together with the condition means and significance levels.
The one-sided paired $t$-test results indicate that the With-feedback condition significantly outperformed the Without-feedback condition in Q3 Attentiveness ($5.29$ vs.\ $4.80$, $p = .018$), Q6 Realness ($4.77$ vs.\ $4.29$, $p = .006$), and Q8 Co-presence ($4.60$ vs.\ $3.94$, $p = .002$).
It also showed marginally significant improvements in Q4 Closeness ($p = .081$) and Q5 Engagement ($p = .089$), while the remaining items showed no significant advantage for the With-feedback condition.
These results suggest that the AI clone's backchanneling and nodding behaviors enhanced users' perceptions of the avatar's attentiveness, the realness of the interaction, and the sense of co-presence~\cite{oh2018systematic}, as well as fostering a greater sense of closeness and engagement with the avatar.
We also verified that the AB/BA counterbalancing did not bias the outcome: an independent-samples comparison of the per-item score differences (With-feedback $-$ Without-feedback) between the two order groups revealed no significant order effect for any of the nine items (all $p > .21$).


\subsection{Interaction Behavior Analysis}

Beyond the subjective ratings, we examined whether the listening behaviors changed the participants' own conversational behavior (Table~\ref{tab:behavior}).
The number of user utterances, the mean utterance length, and the dialogue duration did not differ significantly between the two conditions (paired $t$-tests, all $p > .19$).
This suggests that the improvements in subjective evaluation reflect changes in the users' perception of the interaction rather than changes in their own overt dialogue behavior.
On average, the system produced 27.3 backchannels and 21.8 nods per dialogue in the With-feedback condition; Figure~\ref{fig:freqdist} shows how these counts were distributed across participants, both as totals per dialogue and normalized per user utterance.

As an exploratory analysis, we investigated how the amount and frequency of system feedback shaped the subjective ratings.
Motivated by prior works on the optimal amount of non-verbal behaviors~\cite{poppe2011backchannels,sebo2020influence}, we fitted a quadratic regression to the rating gains (With-feedback $-$ Without-feedback) for Q3, Q6, and Q8 (Figure~\ref{fig:quad}).
While the addition of feedback generally improved ratings, simply providing more was not necessarily better.
Specifically, Attentiveness (Q3) showed a significant inverted-U relationship with the number of backchannels (quadratic term $p = .009$), peaking at an intermediate amount.
Other items and nodding showed no significant curvilinear trends.

These results suggest that the optimal amount of listening feedback is not uniform.
In the context of AI clones, this highlights a crucial future direction: developing adaptive models that optimize listening behaviors by balancing the original person's authentic listening style with the specific interacting user's dynamics.

\section{Conclusion}

As a late-breaking result, we showed that adding verbal backchannels and head nodding to an AI clone increases the perceived attentiveness of the avatar, the sense of talking with the real person, and the feeling of co-presence.
This indicates that the fidelity of an AI clone is shaped not only by its voice and response content but also by how it listens.

Given these preliminary results, several directions remain for future work.
First, we plan to build personalized models of backchanneling and nodding so that an AI clone reproduces the target person's own listening style, rather than a generic one.
Second, the present study cloned a single person; we will evaluate the approach across multiple target individuals to test its generality~\cite{kasahara2026clone}.
Third, we enabled backchannels and nodding together, so future experiments should decompose the two modalities to clarify their individual contributions.
Lastly, since the current experiment was carried out with Japanese subjects, the similar trend needs to be confirmed in other languages and cultures, using multi-lingual models~\cite{inoue2024multilingual,inoue2026multilingual}.

\section*{Safe and Responsible Innovation Statement}

Our AI clones raise risks of impersonation and deception, which are amplified by the increased realness of listening behaviors.
We mitigate these by cloning only a consenting individual (the first author) strictly for this study and explicitly informing participants they are interacting with an AI.
Participant data was consensually collected and anonymized.
Responsible deployment requires the cloned person's explicit consent, transparent disclosure to users, and robust safeguards against misuse.

\begin{acks}
This work was supported by JST PRESTO (JPMJPR24I4, JPMJPR23I4), JST BOOST (JPMJBY24A7), and JST Moonshot R\&D (JPMJPS2011).
\end{acks}

{
\bibliographystyle{ACM-Reference-Format}
\bibliography{reference}
}

\end{document}